 \documentclass[final,5p,times,twocolumn]{elsarticle}

\usepackage{amssymb}
\usepackage{amsmath}
\usepackage{graphicx}
\usepackage{dcolumn}
\usepackage{bm}
\usepackage{hyperref}
\usepackage[mathlines]{lineno}
\usepackage{xcolor} 
\usepackage{amsthm}

\usepackage{mathtools}

\begin{document}

\begin{frontmatter}



\title{Point Defects Limited Carrier Mobility in Janus MoSSe monolayer}

\author[HCMUS,VNU]{Nguyen Tran Gia Bao}
\author[HCMUS,VNU]{Ton Nu Quynh Trang}
\author[INOMAR,Health,VNU]{Phan Bach Thang}
\author[HCMUT,VNU]{Nam Thoai}
\author[HCMUS,VNU]{Vu Thi Hanh Thu\corref{cor1}}
\ead{vththu@hcmus.edu.vn}
\author[NTU,HCMUS,VNU]{Nguyen Tuan Hung\corref{cor2}}
\ead{nguyenth@ntu.edu.tw}
\cortext[cor1]{Corresponding author.}
\cortext[cor2]{Corresponding author.}

\affiliation[HCMUS]{organization={Faculty of Physics and Physics Engineering, University of Science},
            city={Ho Chi Minh City},
            postcode={700000}, 
            country={Viet Nam}}

\affiliation[INOMAR]{
            organization={Advanced Materials Technology Institute Vietnam National University Ho Chi Minh City (formerly affiliated with Center for Innovative Materials and Architectures)},
            city={Ho Chi Minh City},
            postcode={700000}, 
            country={Viet Nam}}
            
\affiliation[Health]{
            organization={University of Health Sciences (UHS), Viet Nam National University Ho Chi Minh City},
            city={Ho Chi Minh City},
            postcode={700000}, 
            country={Viet Nam}}
            
\affiliation[HCMUT]{
            organization={Ho Chi Minh City University of Technology},
            city={Ho Chi Minh City},
            postcode={700000}, 
            country={Viet Nam}}
\affiliation[VNU]{
            organization={Viet Nam National University Ho Chi Minh City},
            city={Ho Chi Minh City},
            postcode={700000}, 
            country={Viet Nam}}
\affiliation[NTU]{
            organization={Department of Materials Science and Engineering, National Taiwan University},
            city={Taipei},
            postcode={10617}, 
            country={Taiwan}}

\begin{abstract}
Point defects, often formed during the growth of Janus MoSSe, act as built-in scatterers and affect carrier transport in electronic devices based on Janus MoSSe. In this study, we employ first-principles calculations to investigate the impact of common defects, such as sulfur vacancies, selenium vacancies, and chalcogen substitutions, on electron transport, and compare their influence with that of mobility limited by phonons. Here, we define the saturation defect concentration ($C_{\mathrm{sat}}$) as the highest defect density that still allows the total mobility to remain within 90\% of the phonon-limited value, providing a direct measure of how many defects a device can tolerate. Based on $C_{\mathrm{sat}}$, we find a clear ranking of defect impact: selenium substituting for sulfur is relatively tolerant, with $C_{\mathrm{sat}}\approx2.07\times10^{-4}$, while selenium vacancies are the most sensitive, with $C_{\mathrm{sat}}\approx3.65\times10^{-5}$. 
Our $C_{\mathrm{sat}}$ benchmarks and defect hierarchy provide quantitative, materials-specific design rules that can guide the fabrication of high-mobility field-effect transistors, electronic devices, and sensors based on Janus MoSSe.
\end{abstract}




\end{frontmatter}



\section{Introduction}
Janus transition metal dichalcogenides (TMDCs) are emerging two-dimensional materials characterized by their unique asymmetrical structure, achieved by differing chalcogen atoms on each side of the metal layer \cite{Bao2025-gv, Hung2017-xi, Hung2023-ot}. This structural asymmetry breaks the out-of-plane mirror symmetry, giving rise to novel properties such as Rashba spin splitting, strong vertical piezoelectricity, and extended exciton and carrier lifetimes \cite{Schmeink2024-rm,Chiu2024-iw,Hu2018-ye}. Due to these intriguing attributes, Janus TMDCs hold significant promise for applications across multiple technological domains, including nanoelectronics, spintronics, optoelectronics, sensors, and catalysis \cite{Zhang2017-ie,Wei2022-ha,Van-Thanh2023-ws,Bao2025-gv,Hung2023-ot,CHAURASIYA2019204}. Particularly, their intrinsic electric field enhances photocatalytic efficiency and gas-sensing capabilities, while their unique electronic structure facilitates efficient hydrogen evolution reactions, highlighting their potential in energy conversion and storage technologies.

Recently, Janus MoSSe and WSSe monolayers have been successfully synthesized through various advanced techniques \cite{Hung2023-ot,Schmeink2024-rm,Trivedi2020-jl,Harris2023-ac}. One common approach is chemical vapor deposition (CVD), where monolayers like MoSe$_2$ or MoS$_2$ are treated to selectively replace the top chalcogen atoms, forming asymmetrical Janus structures. Another technique is pulsed laser deposition (PLD), where clusters of atoms (e.g., Se clusters) are introduced onto an existing TMD monolayer to selectively replace the upper atomic layer \cite{Harris2023-ac}. Additionally, selective epitaxial atomic replacement (SEAR) has enabled the synthesis of Janus TMDCs and their heterostructures even at room temperature by selectively removing and substituting chalcogen atoms \cite{Trivedi2020-jl}.

During the synthesis of Janus TMDCs, a variety of point and extended defects can arise. For example, chalcogen vacancies (missing S/Se atoms) often form due to incomplete substitution or evaporation of the chalcogen species during CVD or PLD growth. Antisite defects, in which a chalcogen atom occupies the metal site or vice versa, can occur if the substitution process is partially selective \cite{Schmeink2024-rm, Xiao2024-dz}. These point defects introduce localized scattering or trapping centers for charge carriers, thereby modifying electron transport and altering the electronic, optical, and catalytic properties of the resulting Janus monolayers \cite{Xiao2024-dz,Lu2019-ch,Spetzler2023-vf}. Therefore, understanding the interactions between the electron and point defects is essential for accurately predicting and optimizing electronic performance and carrier mobility in Janus TMDC-based devices.


\citet{Lu2019-ch} introduced an efficient framework for computing electron–defect (e–d) interactions that delivers atomic-level accuracy while remaining both computationally affordable and systematically convergent. In contrast to conventional all-supercell schemes—which require large supercells (typically containing hundreds of atoms) to obtain electron wavefunctions and defect perturbation potentials—those traditional approaches incur significant computational and storage costs for handling large-scale wavefunctions, evaluating e–d matrix elements, and enforcing strict convergence of relaxation times with respect to supercell size and Brillouin-zone sampling. More recently, a simplified approximation implemented in the EPW code assumes a uniform, random distribution of ionized impurities and neglects neutral impurity (short-range) scattering \cite{Leveillee2023-vu, Rosul2022-jd, Settipalli2022-gc}. By employing an analytic monopole impurity potential together with Kohn–Luttinger ensemble averaging, this technique completely avoids explicit defect-supercell wavefunctions and vastly accelerates computation. However, it cannot capture how variations in defect species alter the electron–defect scattering and the e-d limited mobility. Moreover, \citet{Xiao2024-dz} employed a similar approach to the \citet{Lu2019-ch}'s approaches in the EDI code to accurately compute electron–defect scattering and the resulting carrier mobilities for four common TMDC semiconductors, demonstrating significant differences in mobility for different defect types within the same material. Therefore, the type of defect in TMDCs has a pronounced impact on carrier mobility.

In this paper, we employ the framework developed by \citet{Lu2019-ch} within the PERTURBO software to investigate electron–defect (e–d) interaction scattering phenomena in the MoSSe Janus monolayer. We focus on four intrinsic neutral point-defects, including S- and Se-vacancies, and S- and Se-substitutions, which often occur during MoSSe synthesis, as well as additional neutral impurities (O and Te) introduced at chalcogenide sites. We also calculate the electron–phonon interaction to comprehensively study the carrier transport properties of the MoSSe Janus monolayer in the presence of point defects. 
Finally, we identify the critical defect concentrations at which defect-limited and phonon-limited mobilities intersect, thereby establishing targets for experimental optimization.

\section{Computational details}
In this work, we calculate the electron–defect and electron–phonon interactions in the Janus MoSSe monolayer using first-principles density functional theory as implemented in the Quantum ESPRESSO package \cite{QE-2009,QE-2017}. We employ optimized norm-conserving Vanderbilt (ONCV) PseudoDojo pseudopotentials with the Perdew–Burke–Ernzerhof (PBE) exchange–correlation functional \cite{vanSetten2018, Jollet2014}. These choices are required for the electron–defect interaction calculation because the Kleinman–Bylander form in Eq. \eqref{eq:KBform}, which splits the pseudopotential into a local term $V_L(\bm{r)}$ and a nonlocal term $\hat{V}_{NL}$ in the Eq. \eqref{eq:KBform}, allows us to derive the two independent contributions to the electron–defect matrix element \cite{Lu2019-ch, Lu2020-cv,PhysRevLett.48.1425}. 

\begin{figure*}[ht]
    \centering
    \includegraphics[width=0.8\linewidth]{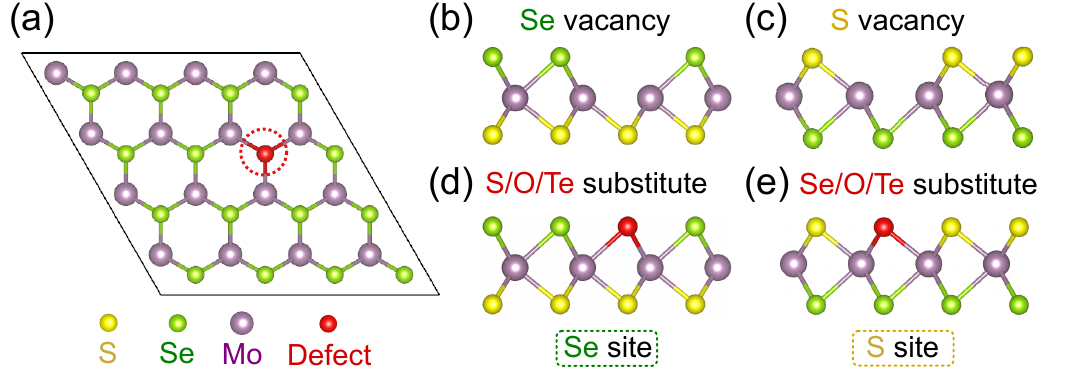}
    \caption{\label{fig:atom}Atomic structures of defect configurations in a Janus MoSSe monolayer. (a) Top view of the pristine $4 \times 4\times 1$ structure, with a representative defect position highlighted by a red sphere and dashed circle, (b) Se vacancy, (c) S vacancy, (d) S/O/Te atom substituting at the Se vacancy site, and (e) Se/O/Te atom substituting at the S vacancy site.}
\end{figure*}

The Janus MoSSe monolayers are modeled using a vacuum spacing large enough (30 $\AA$) to suppress spurious interlayer interactions. Both the primitive cell and all defect-containing supercells are fully relaxed using the Broyden–Fletcher–Goldfarb–Shanno (BFGS) algorithm, with convergence criteria set to $\Delta E < 1.0\times10^{-6}\,\mathrm{Ry}$ for the total energy and $\|\mathbf{F}\| < 1.0\times10^{-5}\,\mathrm{Ry/Bohr}$ for the Hellmann-Feynman forces~\cite{hung2022quantum}. Electronic self-consistency is achieved by a total-energy convergence threshold of $1.0\times10^{-12}\,\mathrm{Ry}$, and a kinetic-energy cutoff of 80 Ry is applied for plane wave kinetic energy. A $12\times12\times1$ Monkhorst–Pack $\mathbf{k}$-point grid is selected for the primitive cell, whereas a $1\times1\times1$ grid is used for the $4\times4\times1$ supercell during both the structural optimization and the ground‐state self‐consistent (SCF) calculations. To achieve dense Brillouin-zone sampling at minimal computational cost, we employed the maximally‐localized Wannier function interpolation scheme as implemented in Wannier90 \cite{Pizzi_2020,PhysRevB.65.035109}. Following a SCF calculation, a non‐self‐consistent field (NSCF) run is performed on a uniform $24\times24\times1$ $\mathbf{k}$‐point grid to obtain converged Bloch eigenstates.

To optimize computation time and focus on which defect type most significantly affects mobility via the electron–defect interaction, we chose a $4\times4\times1$ supercell of the Janus MoSSe monolayer containing a single defect at the center of the supercell, as shown in Fig.~\ref{fig:atom} (a). This supercell size is chosen based on benchmarks in the e–d previous literature showing that neutral-defect perturbations are short-ranged and that moderate supercells already capture most of the signal \cite{Lu2019-ch}. Given the low defect concentration, we approximate the electronic bands of the defect supercell by those of the pristine crystal and evaluate electron–defect scattering with \citet{Lu2019-ch}'s formalism. We treat defects as neutral and non-interacting and work in lowest-order Born perturbation theory, under which the scattering rate for a Bloch state $T^{\pm,\rm d}_{n\bm{k},\,n'\bm{k}'}$ in the Eq. \eqref{eq:T-ed} scales linearly with the dimensionless atomic defect concentration $C_d$ \cite{Lu2019-ch}. Accordingly, we show e–d quantities per unit $C_d$ and rescale linearly to any target concentration at post-processing, provided scattering events remain independent over the range considered.

In defect-interaction calculation, the e–d perturbation potential $\Delta V_{\rm d}$ in Eq. \eqref{eq:M_def} is obtained as the difference of the Kohn–Sham potentials of defect‐containing and pristine supercells. Matrix elements are split into local and nonlocal contributions and computed using only the primitive‐cell wavefunctions, with the perturbation potential Fourier‐transformed and interpolated on a moderate grid of transferred momenta $\mathbf{q}$. The procedure for computing the e-d matrix element is described in the \ref{sec:A3}. The e-d scattering rates are calculated in the lowest‐Born approximation, summing over final states on ultra‐fine BZ grids (up to $150 \times150 \times 1$ points) for the mobility calculations. Defect‑limited carrier mobilities are then obtained by solving the linearized Boltzmann transport equation using the relaxation time approximation approach (RTA), with a Gaussian broadening of 2 meV employed to approximate the Dirac delta function in the scattering integrals (Eq. \eqref{eq:T-ed}). All e–d workflows—including matrix‐element generation, convergence studies in supercell size, and $\mathbf{k}$-grid density, as well as subsequent relaxation‐time and mobility calculations are calculated by using the PERTURBO code \cite{Lu2019-ch, Lu2020-cv}.

Electron–phonon scattering rates are also calculated with the PERTURBO code \cite{Zhou2021-hc}. The required inputs are the electron eigenvalues and band velocities from the NSCF calculation, maximally localized Wannier functions, and the dynamical matrices and first-order perturbation potentials obtained from DFPT \cite{Pizzi_2020,PhysRevB.65.035109}. To remove spurious interactions between periodic images in the out-of-plane direction, all slab calculations employed a two-dimensional Coulomb cutoff \cite{PhysRevB.96.075448}, which truncates the Coulomb interaction along \(z\) for structures periodic in the \(xy\) plane. Dynamical matrices and phonon perturbations are first computed on a coarse \(6\times6\times1\) \(\mathbf{q}\)-point mesh~\cite{RevModPhys.73.515} and then they are interpolated onto dense \(150\times150\times1\) \(k\)- and \(q\)-grids using the Wannier representation, enabling accurate evaluation of electron–phonon scattering. A Gaussian broadening of \(5~\mathrm{meV}\) is used to approximate the energy-conserving Dirac delta functions in the scattering integrals (Eq.~\eqref{eq:T_ph_branch}).

Finally, to combine the effects of electron–defect (e–d) and electron–phonon (e–ph) scattering, we adopt the Matthiessen rule \cite{PhysRevMaterials.6.L010801}, which assumes each scattering channel contributes independently to the total relaxation rate. The total mobility, $\mu_{\rm total}$, is defined as follows:
\begin{equation}
    \mu_{\rm total}^{-1} = \mu_{\rm e-d}^{-1} + \mu_{\rm e-ph}^{-1},
\end{equation}
where $\mu_{\rm e-d}$ and $\mu_{\rm e-ph}$ are the mobility limited by the point-defect and phonon, respectively.

\section{Result and Discussion}

\subsection{The defect types, geometric, electronic structures, and phonon dispersion}

\begin{figure}
    \centering
    \includegraphics[width=0.9\linewidth]{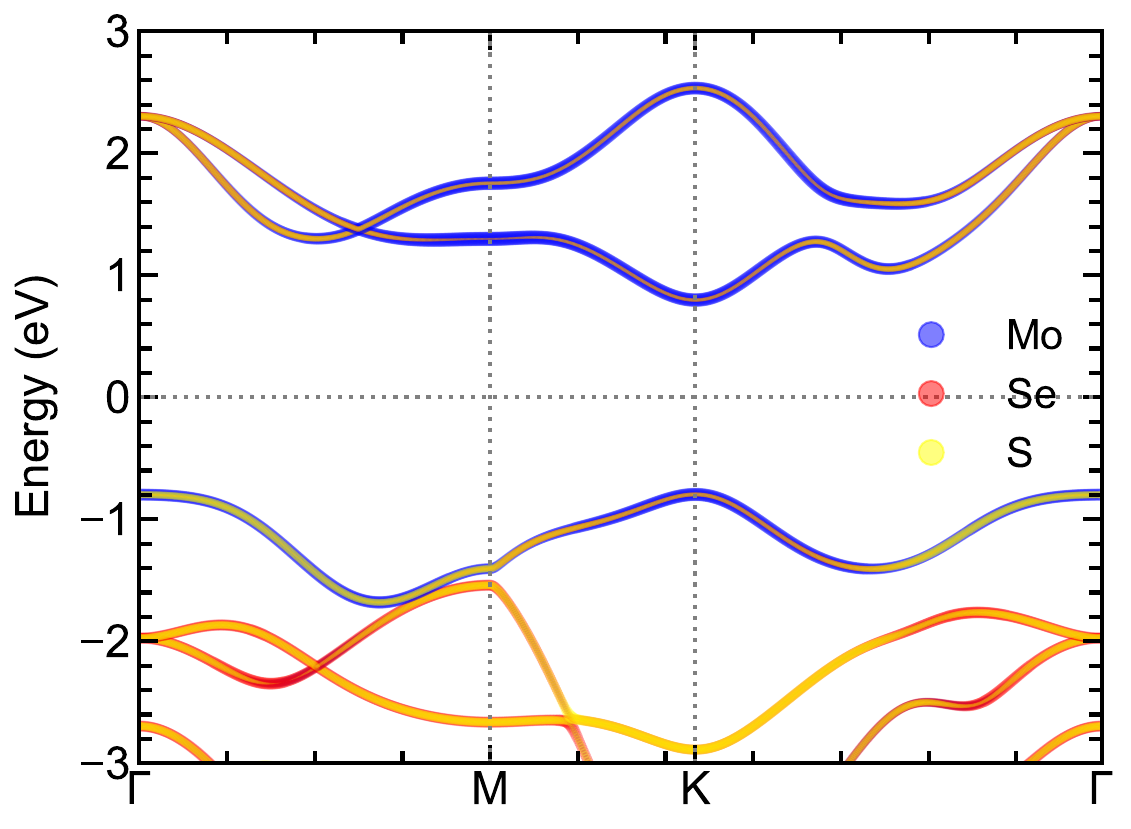}
    \caption{\label{fig:bandstrcuture}Orbital‐projected band structure of the MoSSe monolayer along the high‐symmetry path. Circle areas scale with the projection weight onto Mo (blue), Se (red), and S (yellow) orbitals.}
\end{figure}

In Fig.~\ref{fig:atom}, we show several types of point-defects in the MoSSe monolayer, including Se and S vacancies and S/O/Te and Se/O/Te substitutes. In the experiment, the MoSSe monolayer has been synthesized by first growing MoS$_2$ or MoSe$_2$ via CVD and then selectively substituting the top chalcogen layer: S atoms in MoS$_2$ are partially replaced by Se vapor (or vice versa) under controlled conditions \cite{Hung2023-ot, LAKSHMY2024100204}. An alternative \textit{in situ} route sequentially activates Mo and S precursors to form MoS$_2$, introduces surface S vacancies through temperature or H$_2$ plasma treatment, and then exposes the layer to Se to occupy those vacancies. Although both approaches yield high‐quality Janus films, they invariably introduce atomic‐scale imperfections during the substitution step \cite{LAKSHMY2024100204}. Point vacancies can arise when the incoming chalcogenide species fail to occupy all lattice sites, and residual original atoms may persist if removal is incomplete.

To reflect defects commonly present after synthesis, we first analyze two intrinsic point defects that arise when converting MoS\(_2\) and MoSe\(_2\) precursors into Janus MoSSe: (i) chalcogen (S or Se) vacancies and (ii) incomplete (S or Se) substitution at the top chalcogen layer~\cite{LAKSHMY2024100204}.
Beyond these native defects, intentional incorporation of other chalcogens provides a handle to tune the defect-limited carrier mobility. In gas-sensing contexts, \citet{CHAURASIYA2019204} reported that exposure of MoSSe to NO\(_2\) can promote oxygen substitution at surface chalcogen sites, which in turn modifies mobility and electrical conductivity. Motivated by these observations, we also model two extrinsic defect species corresponding to Te and O substitution at the Janus surface, as shown in Figs.~\ref{fig:atom} (d) and (e).
In total, our framework examines eight defect configurations, enabling a systematic assessment of how both synthesis-induced and intentionally introduced point defects govern electron–defect scattering in the MoSSe monolayer.

The optimized structures of the MoSSe monolayer show the lattice constant $a = 3.25$ $\AA$, which is consistent with the experimental value ($3.22\pm0.01$ $\AA$~\cite{picker2025atomic}). In the Fig. \ref{fig:bandstrcuture}, we plot the Wannier interpolated band structure. The Wannier functions (WFs) are constructed by projecting onto a reduced set of atomic‐like trial orbitals: the Mo $d_{xy}$, $d_{x^2-y^2}$, and $d_{z^2}$ orbitals, and the chalcogen (S and Se) $p_x$, $p_y$, and $p_z$ orbitals, yielding nine WFs per cell \cite{PhysRevB.104.L161108}. 
The calculated band structure of Janus MoSSe indicates that the conduction band minimum (CBM) and valence band maximum (VBM) occur at the K and $\Gamma$ points, exhibits an indirect band gap of $1.59$ eV, in agreement with experimental reports \cite{Hung2023-ot}. We note that the current version of PERTURBO does not support spin–orbit coupling (SOC) in e-d calculations. Thus, we use scalar‐relativistic PBE eigenvalues (without SOC) to extract the electronic states for both e-d and e-ph calculations.

\begin{table*}[t]
\centering
\setlength{\tabcolsep}{8pt}
\renewcommand{\arraystretch}{1.1}
\caption{\label{tab:defect_mobility}
Defect-limited (e-d) and total (e-d + e-ph) drift mobilities for monolayer MoSSe.
The e-d mobilities are evaluated at a representative defect concentration
$C_d=10^{-3}$ (1000\,ppm). Total mobilities combine e-d scattering with intrinsic
phonon scattering by using Matthiessen’s rule using $\mu_{\text{e-ph}}^{e}=59.34$ and
$\mu_{\text{e-ph}}^{h}=8.03$~cm$^{2}$V$^{-1}$s$^{-1}$.}
\begin{tabular*}{\textwidth}{@{\extracolsep{\fill}}l c r r r r}
\hline
\multicolumn{2}{c}{} & \multicolumn{2}{c}{e-d mobility (cm$^{2}$/V\,s)} & \multicolumn{2}{c}{e-d + e-ph (cm$^{2}$/V\,s)} \\
Defect type & Site & Electron & Hole & Electron & Hole \\
\hline
S vacancy                 & S  &  47.86 & 11.69 & 26.49 &  4.76 \\
Se vacancy                & Se &  19.99 &  1.91 & 14.95 &  1.55 \\
Se substitute (Se-S)      & S  & 117.29 & 95.00 & 39.40 &  7.40 \\
S substitute (S-Se)       & Se &  84.38 &108.51 & 34.84 &  7.48 \\
O substitute (O-S)        & S  &   9.14 &  5.54 &  7.92 &  3.28 \\
Te substitute (Te-Se)     & Se &  25.04 & 30.31 & 17.61 &  6.35 \\
O substitute (O-Se)       & Se &   4.46 &  7.09 &  4.15 &  3.77 \\
Te substitute (Te-S)      & S  &  13.51 & 18.85 & 11.00 &  5.63 \\
\hline
\end{tabular*}
\end{table*}

\subsection{Defect-limited carrier mobility}
At room temperature $T=300$ K, we calculate defect-limited mobilities for eight neutral point-defect configurations with the dimensionless atomic defect concentration of \(C_d=10^{-3}\) (see Eq.~\eqref{eq:T-ed}) (one defect per \(10^{3}\) host atoms, which equivalent to $3.28\times10^{9}$ cm$^{-2}$).
Mobilities are obtained by solving the linearized Boltzmann transport equation in the relaxation-time approximation (RTA) using state-resolved e–d rates (see~\ref{sec:A1}).
It is noted that in our framework, point defects are assumed not to modify the pristine band structure; consequently, the relation between carrier concentration and chemical potential is identical across all configurations for the e–d and e-ph limited calculations \cite{Lu2020-cv}.

As shown in Table~\ref{tab:defect_mobility}, among the native defects that may form during synthesis, the Se substitution at the S site yields the highest defect-limited electron mobility at the CBM, reaching \(117.29~\mathrm{cm}^2\mathrm{V}^{-1}\mathrm{s}^{-1}\). In contrast, the Se vacancy introduces the strongest scattering effect, resulting in the lowest electron mobility of 
\(19.99~\mathrm{cm}^2\mathrm{V}^{-1}\mathrm{s}^{-1}\). For the hole mobility: the S substitution at the Se site provides the largest value (\(108.51~\mathrm{cm}^2\mathrm{V}^{-1}\mathrm{s}^{-1}\)), whereas the Se vacancy leads to the most severe reduction, yielding only \(1.91~\mathrm{cm}^2\mathrm{V}^{-1}\mathrm{s}^{-1}\).  

When intrinsic phonon scattering is also included, the same hierarchy persists, but the absolute values decrease due to the combined scattering channels. For electrons, the Se substitution at the S site retains the highest total mobility (\(39.40~\mathrm{cm}^2\mathrm{V}^{-1}\mathrm{s}^{-1}\)), 
while the Se vacancy remains the lowest at \(14.95~\mathrm{cm}^2\mathrm{V}^{-1}\mathrm{s}^{-1}\). 
Similarly, for holes, the Se substitution at the S site achieves the maximum value (\(7.40~\mathrm{cm}^2\mathrm{V}^{-1}\mathrm{s}^{-1}\)), whereas the Se vacancy reduces the mobility drastically to only \(1.55~\mathrm{cm}^2\mathrm{V}^{-1}\mathrm{s}^{-1}\). This consistent trend highlights the detrimental role of chalcogen vacancies in limiting carrier transport, in contrast to chalcogen substitutions, which are relatively less disruptive.  

In addition to native defects, the intentional incorporation of other chalcogens further illustrates the sensitivity of carrier mobility to substitutional chemistry. Oxygen substitution at either the S or Se site leads to a pronounced degradation of both electron and hole transport, with total mobilities suppressed to the range of \(3\text{-}8~\mathrm{cm}^2\mathrm{V}^{-1}\mathrm{s}^{-1}\). By comparison, tellurium substitution is less detrimental: Te occupying the Se site results in moderate mobilities (\(17.61~\mathrm{cm}^2\mathrm{V}^{-1}\mathrm{s}^{-1}\) for electrons and \(6.35~\mathrm{cm}^2\mathrm{V}^{-1}\mathrm{s}^{-1}\) for holes), while Te at the S site yields slightly lower values (\(11.00~\mathrm{cm}^2\mathrm{V}^{-1}\mathrm{s}^{-1}\) and \(5.63~\mathrm{cm}^2\mathrm{V}^{-1}\mathrm{s}^{-1}\), respectively). These results suggest that while Se substitution enhances transport, the incorporation of smaller (O) or larger (Te) chalcogens introduces stronger scattering, thereby reducing mobility.  

\begin{figure}
    \centering
    \includegraphics[width=0.85\linewidth]{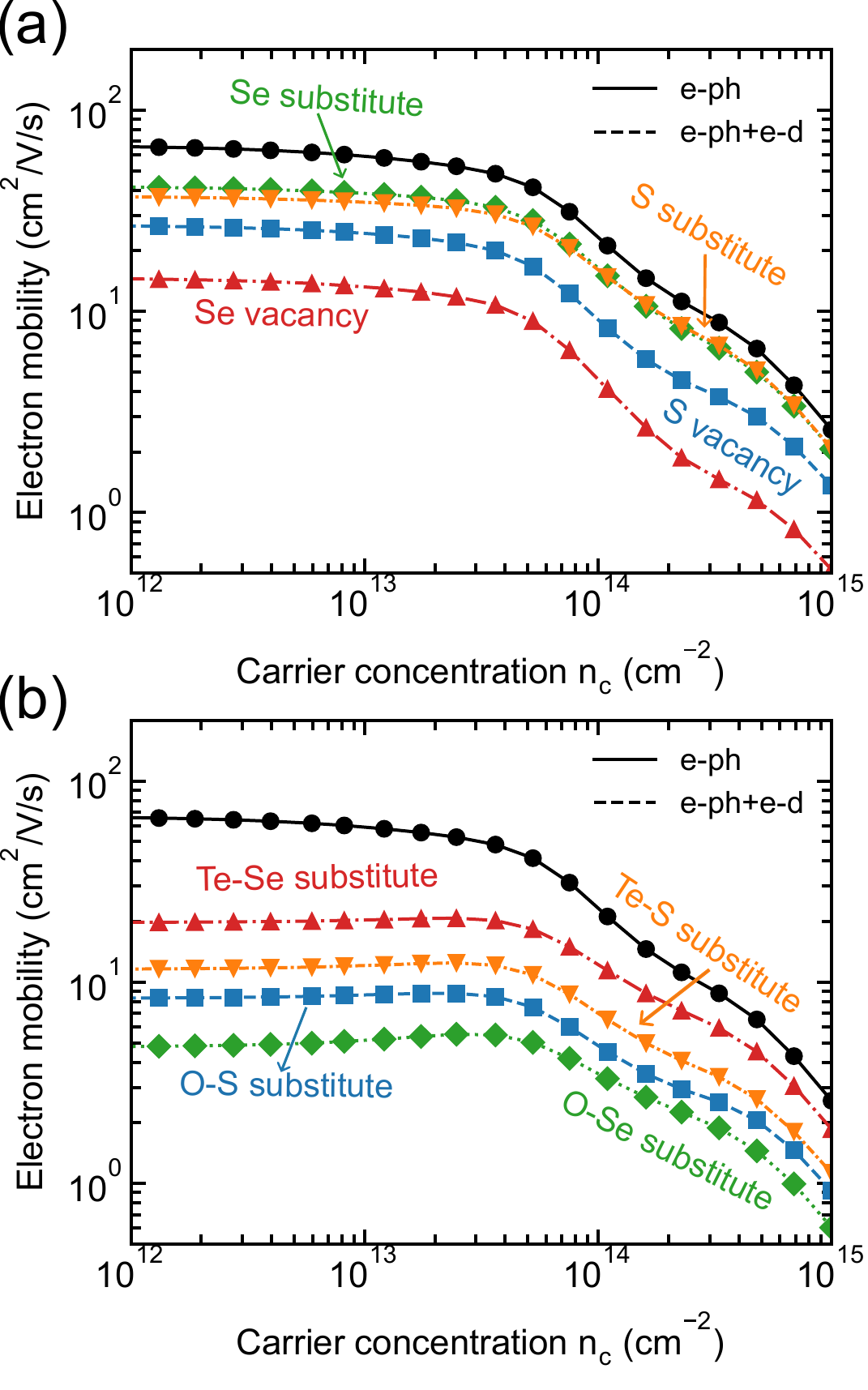}
    \caption{\label{fig:carrier-mob}Electron mobilities in defected MoSSe as functions of carrier concentration at 300 K. Calculations include phonon‐limited mobility (solid line) and combined defect‐ and phonon‐limited mobility (dashed line).}
\end{figure}

In Fig.~\ref{fig:carrier-mob}, we show the phonon-limited (e-ph) and the combined defect- and phonon-limited (e-d + e-ph) mobilities as functions of carrier concentration at 300~K. For a complete comparison including both electron and hole carriers, the corresponding results are provided in Figs.~S1 and S2 of the Supporting Information. As shown in Fig.~\ref{fig:carrier-mob}, the separation between the solid (e-ph) and dashed (e-ph + e-d) curves directly reflects the relative importance of defect scattering. When the dashed line nearly overlaps with the solid line, the total mobility is dominated by e-ph interactions, whereas a larger separation signifies a stronger contribution from e-d scattering. In Fig.~\ref{fig:carrier-mob}(a), the spacing between the two curves remains relatively constant over the range of carrier concentrations, indicating that the relative weight of e-ph and e-d contributions is not strongly dependent on carrier density. Among the native defects, the Se vacancy produces the largest deviation between the dashed and solid curves, demonstrating that its total mobility is strongly suppressed by e-d scattering. In contrast, Se and S substitutions yield curves that remain close to the e-ph baseline, implying that, at the same defect concentration of \(C_d = 10^{-3}\), these substitutional defects introduce only weak scattering and the mobility is largely governed by intrinsic phonon interactions.  

This contrast reflects the strength of the defect-induced perturbation. Vacancy defects, such as the removal of a Se or S atom, eliminate both the local electrostatic potential and the relaxed atomic positions, thereby creating a strong perturbation \(\Delta V_d\) and correspondingly large scattering matrix elements \(M_{n'\mathbf{k},n\mathbf{k}}\) in Eq.~\eqref{eq:M_def}. 
As a result, the total mobility is dramatically reduced. In comparison, substitutional defects replace a chalcogen atom with a chemically similar species, leading to a much smaller \(\Delta V_d\) and weaker scattering. Consequently, the defect-limited mobilities for substitutional cases remain substantially higher. A similar trend is also observed for the hole mobility (see Fig.~S2 in the Supporting Information). In addition to native defects, the effect of intentional substitution by other chalcogen species is plotted in Fig.~\ref{fig:carrier-mob}(b). Both O and Te substitutions lead to a stronger suppression of mobility compared to the intrinsic S and Se substitutions. This reduction arises from the pronounced mismatch in ionic radius and electronegativity between the host chalcogen atoms and the substituents, which produces a larger perturbation potential \(\Delta V_d\) and thus stronger e-d scattering. In particular, the O substitution at either the S or Se site causes a severe mobility degradation, as the smaller ionic size and high electronegativity of the O substitution introduce strong local distortions and enhance carrier scattering. The Te substitution, while less disruptive than the O substitution, still reduces the mobility significantly relative to Se or S substitutions; its larger atomic size generates strain and modifies the local electronic potential, thereby increasing scattering. Overall, the mobility reduction induced by these intentional substitutions exceeds that of as-synthesized intrinsic substitutional defects, underscoring the sensitivity of carrier transport in MoSSe to chemical mismatch at chalcogen sites.  

\begin{figure}[t]
    \centering
    \includegraphics[width=0.85\linewidth]{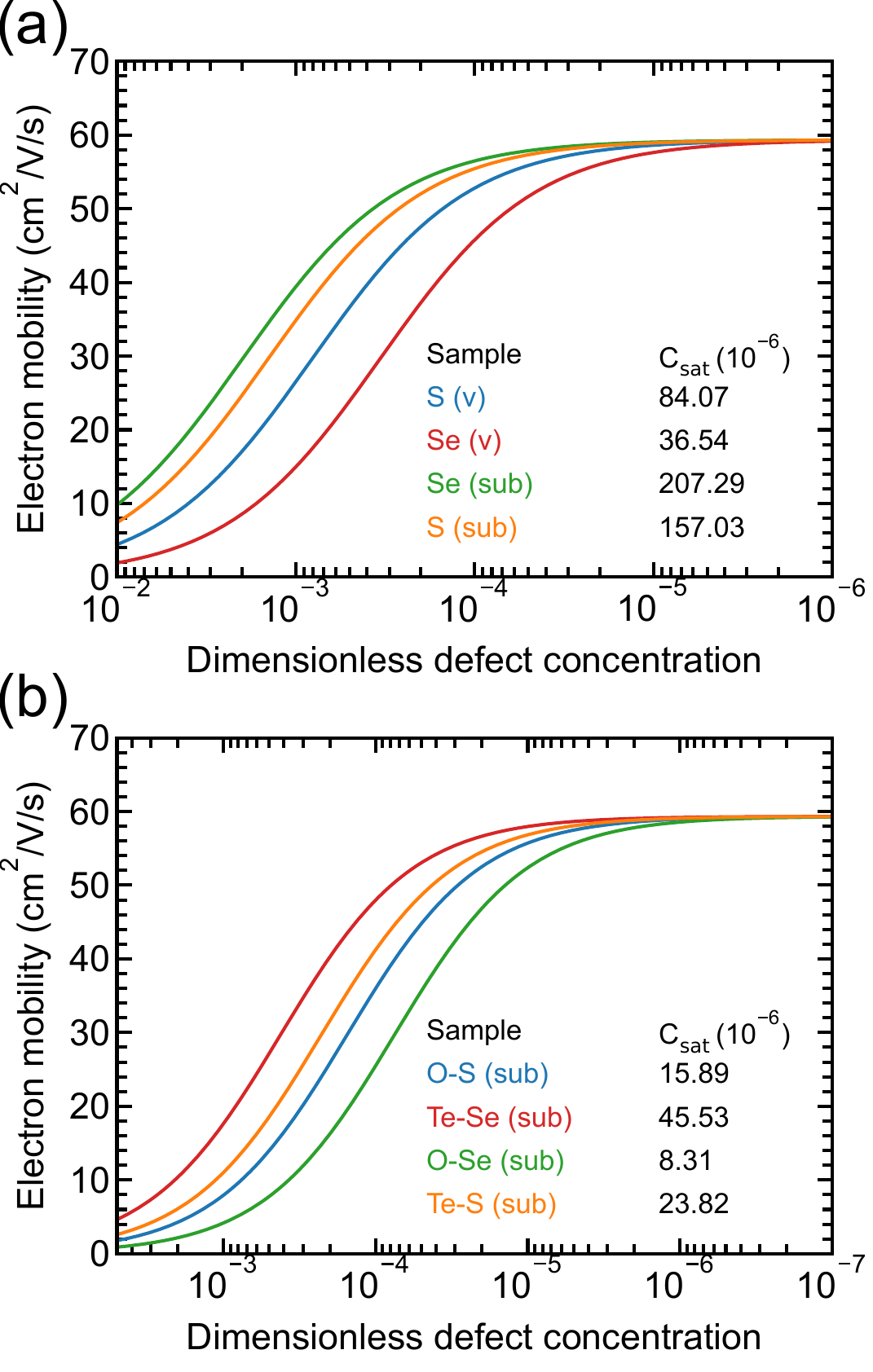}
    \caption{\label{fig:defect-mob}Combined defect and phonon-limited electron mobility in defected MoSSe as a function of defect concentration $C_d$. The phonon-limited mobility is $\mu_{\text{e-ph}} = 59.34$~cm$^{2}$V$^{-1}$s$^{-1}$ at a carrier concentration of $1.2 \times 10^{13}$~cm$^{-2}$.}
\end{figure}

To quantify how defect concentration influences the total mobility, we computed the defect–limited electron mobility \(\mu_{\text{e-d}}(C_{\mathrm{d}})\) over a range of defect fractions and combined it with the phonon–limited mobility at \(T=300~\mathrm{K}\), as shown in Fig.~\ref{fig:defect-mob}. Here, the e-d mobilities are evaluated at a defect concentration of 1 ppm for the reference, using the same carrier concentration as in the electron–phonon interaction case. According to Eq. \ref{eq:T-ed}, the relaxation time is linearly proportional to $C_d$, implying that the mobility is inversely proportional to $C_d$. Therefore, the total mobility can be obtained by using Matthiessen’s rule as follows:
\begin{equation}
    \frac{1}{\mu_{\text{total}}}
    = \frac{1}{\mu_{\text{e-ph}}}
    + \left( \frac{C_d}{1 \times 10^{-6}} \right)
      \frac{1}{\mu_{\text{e-d}\text{ at 1 ppm}}}.
\end{equation}
 
In Fig.~\ref{fig:defect-mob}, all mobilities are evaluated at a carrier concentration of \(n_c = 1.2 \times 10^{13}~\mathrm{cm}^{-2}\). At $T=300$ K, when $C_d$ decreases from \(10^{-2}\) to \(10^{-6}\), \(\mu_{\mathrm{total}}\) increases monotonically and rapidly approaches the phonon–limited value \(\mu_{\text{e-ph}}\approx 59.34~\mathrm{cm}^2\,\mathrm{V}^{-1}\,\mathrm{s}^{-1}\). This dependence of \(\mu_{\mathrm{total}}\) on \(C_{\mathrm{d}}\) provides practical guidance for experimental optimization of growth and processing conditions. To make this connection quantitative, we define the saturation defect concentration, \(C_{\mathrm{sat}}\), as the highest defect density for which the total mobility remains within 90\% of the phonon-limited value, i.e.\ the threshold at which \(\mu_{\mathrm{total}} < 0.9\,\mu_{\text{e-ph}}\). This criterion captures the maximum tolerable defect density before defect scattering produces a significant mobility degradation.

From Fig.~\ref{fig:defect-mob}, we extract \(C_{\mathrm{sat}}\) for each defect species. Among the intrinsic defects Fig. ~\ref{fig:defect-mob}(a), the Se substitution at the S site yields the largest \(C_{\mathrm{sat}}\) (\(207.29\times 10^{-6}\)), indicating that this defect is relatively benign and that the mobility remains phonon-limited even up to high defect densities. By contrast, Se vacancies exhibit the smallest \(C_{\mathrm{sat}}\) (\(36.54\times 10^{-6}\)), reflecting their strong scattering effect and low tolerance in synthesis. S vacancies and S substitutions fall in between, with \(C_{\mathrm{sat}} = 84.07\times 10^{-6}\) and \(157.03\times 10^{-6}\), respectively. For intentional chalcogen substitutions, in Fig.~\ref{fig:defect-mob}(b), a similar hierarchy is observed. The O substitution at the Se site is the most detrimental, with the lowest \(C_{\mathrm{sat}} = 8.31\times 10^{-6}\), showing that even trace oxygen incorporation can significantly suppress the electron mobility. The Te substitution at the Se site, on the other hand, is comparatively less sensitive, with a much higher \(C_{\mathrm{sat}} = 45.53\times 10^{-6}\). O substitution at the S site (\(15.89\times 10^{-6}\)) and Te substitution at the S site (\(23.82\times 10^{-6}\)) lie between these two extremes.

Overall, these results highlight two complementary implications. First, defects with large \(C_{\mathrm{sat}}\), such as Se substitution on S sites, define a wide defect tolerance window and are thus compatible with high-mobility applications in field-effect transistors and photocatalysts. Second, the very low \(C_{\mathrm{sat}}\) of oxygen-related substitutions implies that unintentional O incorporation during growth must be carefully avoided for mobility-critical devices. On the other hand, this strong sensitivity to oxygen makes MoSSe a promising candidate for chemical sensing: even trace levels of O-containing molecules such as NO or NO\(_2\) could measurably reduce mobility, enabling highly responsive multifunctional sensors \cite{CHAURASIYA2019204}.  

\subsection{Temperature-dependent carrier mobility in the presence of point defects}

\begin{figure*}[t]
    \centering
    \includegraphics[width=0.85\linewidth]{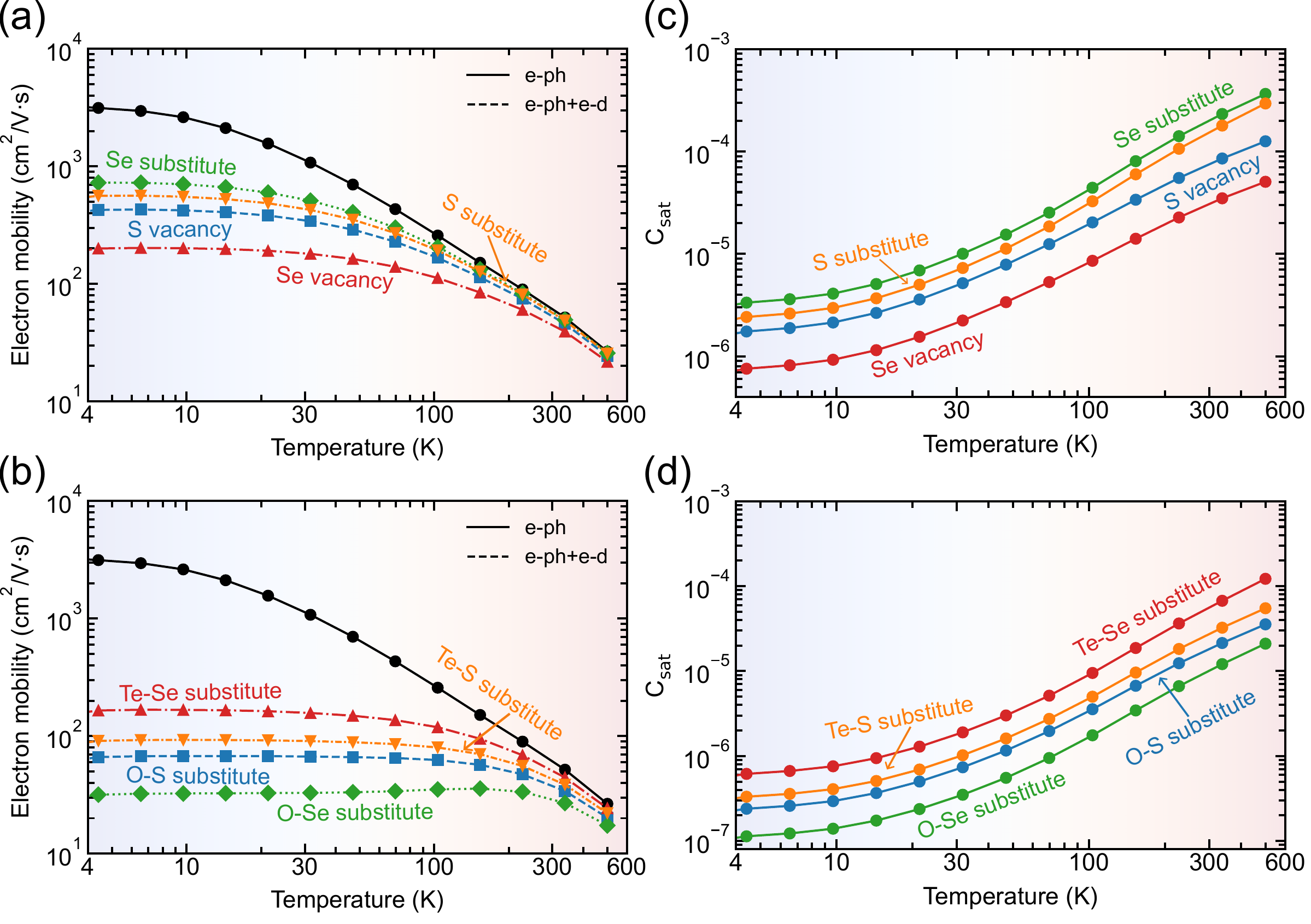}
    \caption{\label{fig:mob-temp} (a)-(b) Electron mobility in defected MoSSe as a function of temperature at a carrier concentration of $1.0 \times 10^{13}$~cm$^{-2}$, including the phonon-limited mobility (solid line) and the combined defect- and phonon-limited mobility (dashed line).}
\end{figure*}

In Fig.~\ref{fig:mob-temp}, we show the temperature dependence of electron mobility in Janus MoSSe, disentangling the roles of phonon scattering and defect scattering. The solid black line corresponds to the phonon-limited mobility, while the dashed colored lines represent the combined effect of phonons and specific defects at a representative concentration of \(C_{\mathrm d}=10^{-4}\) and carrier density \(n=10^{13}\,\mathrm{cm}^{-2}\). As shown in Fig. \ref{fig:mob-temp}(a), intrinsic chalcogen-site defects such as S/Se vacancies and S-Se substitutions introduce distinct suppression of mobility, while Fig. \ref{fig:mob-temp}(b) highlights the stronger impact of hetero-atom substitutions (O and Te at S/Se sites). At cryogenic temperatures (\(T\leq 50\)) K, phonons are frozen out, and the mobility plateaus at nearly constant values determined solely by defect scattering. Vacancies are relatively benign, sustaining higher mobility, whereas substitutional impurities (especially O and Te) strongly depress transport. 

With increasing temperature (\(200 \leq T \leq 600\)) K, the mobility curves for the combined e–ph + e–d scattering can also be described by a power law, \(\mu(T)\propto T^{-\alpha}\). In this regime, the slopes are steeper than those of the pure e–ph case, which indicates that defect scattering still makes a significant contribution to charge transport. A stronger deviation from the phonon-limited reference line, therefore, reflects the enhanced role of defects in suppressing mobility. Temperature-dependent electron and hole carriers for all point defects are shown in Figs. S3 and S4, respectively, of the Supporting Information. 

For the case of defect scattering alone (see Fig.~S5 in the Supporting Information), the behavior differs depending on the type of defect. Vacancies yield nearly constant mobility at low temperatures, followed by a monotonic decrease as temperature increases. By contrast, substitutional impurities display a non-monotonic trend: the mobility initially increases with temperature, reaches a maximum, and then decreases above \(\sim 400~\mathrm{K}\). 
In this range, the data are well described by a power law \(\mu(T)\propto T^{-\alpha}\) for electrons, with best-fit exponents \(\alpha=1.20\) for Se vacancies, and \(\alpha=1.11\) for S vacancies. 


After including e–ph scattering, the behavior above \(200~\mathrm{K}\) becomes phonon-limited: the mobility curves for different defects collapse onto a common temperature dependence set by e–ph processes, \(\mu(T)\propto T^{-\alpha_{\mathrm{ph}}}\), with a defect-independent exponent \(\alpha_{\mathrm{ph}}=1.54\) for electron. At elevated temperatures (\(T\geq 200\) K), all mobility curves converge toward the phonon-limited envelope, indicating that high-temperature charge transport is ultimately dictated by electron–phonon processes, with only modest variations across defect species.

Moreover, to evaluate optimized parameters for experiments at varying temperatures, we calculate how \(C_{\mathrm{sat}}\) evolves with temperature and defect type, as shown in Fig.~\ref{fig:mob-temp}. 
At high temperature, \(C_{\mathrm{sat}}\) is larger than at low temperature. This reflects the fact that, when phonon scattering dominates, defects play a secondary role and higher concentrations can be tolerated without strongly reducing mobility. In contrast, at cryogenic temperatures the phonon contribution is suppressed, so transport is almost entirely defect-limited, making the mobility extremely sensitive to even ppm-level variations in \(C_{\mathrm{d}}\). 

As shown in Fig.~\ref{fig:mob-temp}(c), among the intrinsic defects, Se substitution at the S site exhibits the highest tolerance, with \(C_{\mathrm{sat}}\) increasing from \(3.34\times 10^{-6}\) at 4~K to \(364.40\times 10^{-6}\) at 500~K. By contrast, the Se vacancy is the most detrimental, 
with \(C_{\mathrm{sat}}\) remaining as low as \(0.76\times 10^{-6}\) at 4~K and only reaching 
\(50.53\times 10^{-6}\) at 500~K. S vacancies and S substitutions fall in between, demonstrating intermediate tolerance levels. For intentional substitutions (see Fig.~\ref{fig:mob-temp}(d)), oxygen-related defects impose the strictest constraints. In particular, O substitution at the Se site yields the lowest \(C_{\mathrm{sat}}\), with values ranging from \(1.1\times 10^{-7}\) at 4~K to just \(2.1\times 10^{-5}\) at 500~K, indicating that even trace oxygen incorporation significantly limits mobility. O substitution at the S site is similarly restrictive, though slightly less severe. Tellurium substitutions, on the other hand, are comparatively more benign: Te at the Se site shows the highest tolerance among the intentional cases (\(0.62\times 10^{-6}\) at 4~K and \(1.2\times 10^{-4}\) at 500~K), followed by Te at the S site. 

\subsection{Carrier energy–dependent relaxation time}

\begin{figure}[t]
    \centering
    \includegraphics[width=0.85\linewidth]{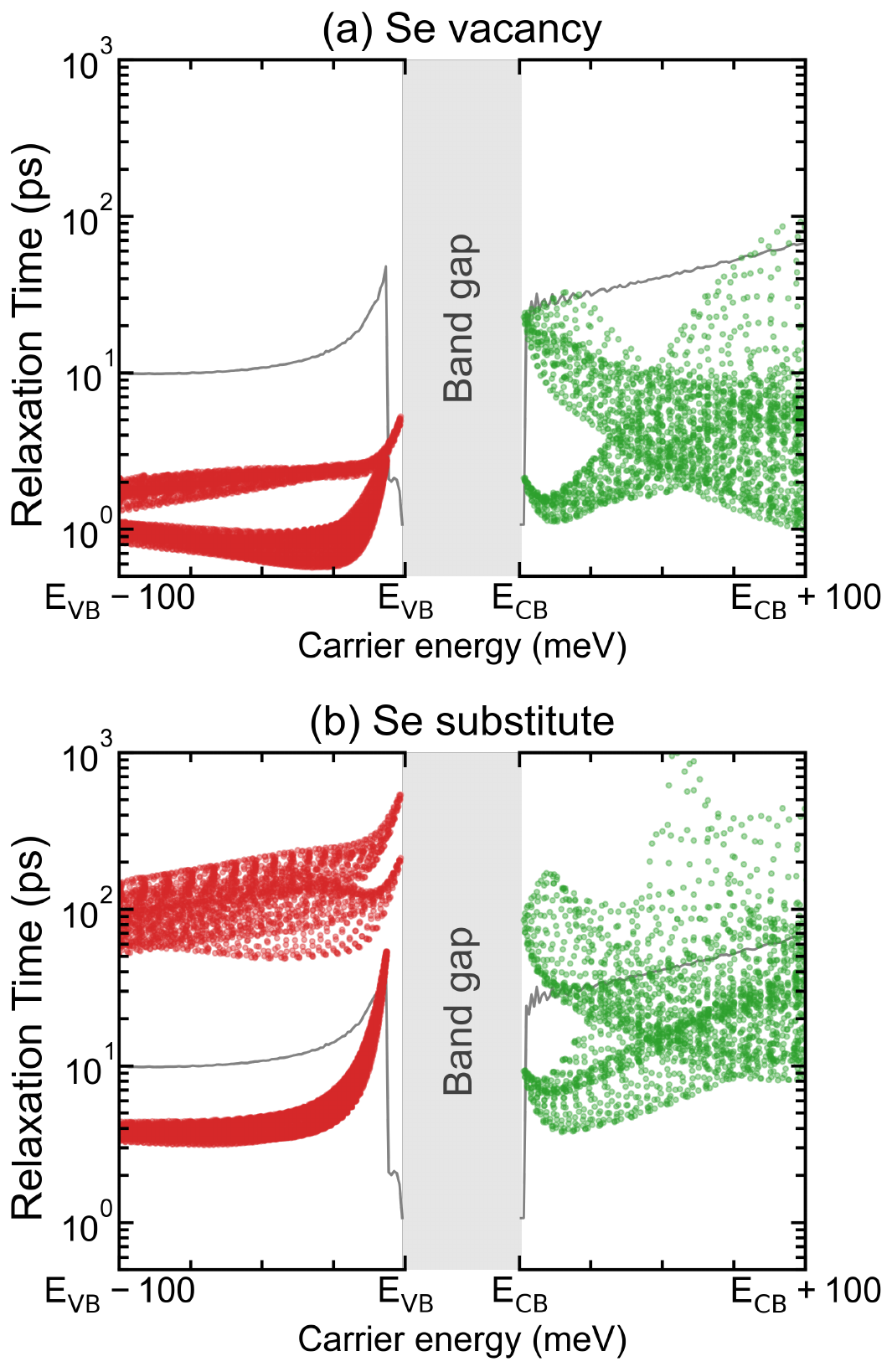}
    \caption{Semi-logarithmic plot of the energy-dependent relaxation time $\tau(E)$ in MoSSe, showing holes (red dots; referenced to $E_{\mathrm{VB}}$) and electrons (green dots; referenced to $E_{\mathrm{CB}}$). Thin gray curves represent the electronic DOS, scaled to arbitrary units for visual comparison with $\tau(E)$.}
    \label{fig:RTs}
\end{figure}

In our earlier discussion, we qualitatively contrasted vacancy and substitutional defects by examining the extent to which each perturbs the electronic states. Here, we quantify this perturbation by calculating the carrier energy–dependent relaxation time, $\tau(E)$, at a defect concentration $C_d = 1$ ppm for both the Se-vacancy and substitution (see Fig. \ref{fig:RTs}). Since the scattering probability at a given energy increases with the availability of final electronic states, $\tau(E)$ is shorter in regions with a large density of states (DOS) and longer in low-DOS regions. Consequently, the band-structure–projected $\tau$ exhibits pronounced minima at energies corresponding to DOS peaks. Comparing the two defect types, the stronger perturbation introduced by a vacancy leads to systematically shorter $\tau$ values across most of the conduction and valence bands, whereas the weaker perturbation of a Se substitution at an S site yields longer relaxation times, particularly in the flatter, low-DOS regions near the band edges. This indicates that Se vacancies degrade carrier transport more severely than Se substitutions. It should be noted that the relaxation times (RTs) for e-d interactions are independent of carrier concentration and temperature, as expressed in Eq. \eqref{eq:T-ed}. Specifically, the RTs are related to the scattering rates via $\tau_{n\mathbf{k},n'\mathbf{k}'} = T_{n\mathbf{k},n'\mathbf{k}'}^{-1}$, which implies $\tau_{n\mathbf{k},n'\mathbf{k}'} \propto C_d^{-1}$. The carrier energy–dependent relaxation time for all point defects is plotted in Fig. S6 in the Supporting Information.

\section{Conclusion}
We quantified how representative neutral point defects govern charge transport in monolayer Janus MoSSe by combining first-principles electron–defect (e-d) scattering with electron–phonon (e-ph) interactions in a Boltzmann-transport equation. Across eight configurations, we observe a robust hierarchy: vacancies—especially V$_\mathrm{Se}$—produce the strongest mobility degradation, whereas isovalent chalcogen substitutions are comparatively benign. Among intrinsic configurations, Se substitution on the S site preserves the highest electron mobility and exhibits the largest tolerance window, with \(C_{\mathrm{sat}}\!\approx\!2.07\times10^{-4}\); by contrast, V$_\mathrm{Se}$ yields the smallest \(C_{\mathrm{sat}}\!\approx\!3.65\times10^{-5}\). Extrinsic chemistry amplifies this contrast: oxygen substitution is particularly detrimental (e.g., Se-site \(C_{\mathrm{sat}}\!\approx\!8.31\times10^{-6}\)), whereas Te substitutions are less severe but still inferior to intrinsic Se/S swaps.

Temperature further stratifies the transport regimes. At cryogenic \(T\), mobilities plateau at defect-limited values, rendering devices exquisitely sensitive to ppm-level \(C_d\). With increasing \(T\), e-ph scattering dominates; the e-d contribution becomes secondary, and \(C_{\mathrm{sat}}\) grows by orders of magnitude (e.g., for Se-S, from \(\sim3.3\times10^{-6}\) at 4\,K to \(\sim3.64\times10^{-4}\) near 500\,K). These trends translate into concrete synthesis and processing targets: suppress chalcogen vacancies, especially V$_\mathrm{Se}$, and rigorously limit oxygen incorporation during growth and post-treatment. Conversely, the pronounced mobility response to trace oxygen also suggests a route to high-responsivity chemical sensing based on MoSSe.


\section*{CRediT authorship contribution statement}
\textbf{Nguyen Tran Gia Bao:} Writing - original draft, Visualization, Data curation, Methodology, Investigation, Formal analysis. \textbf{Ton Nu Quynh Trang:} Methodology, Investigation. \textbf{Phan Bach Thang:} Resources, Funding acquisition. \textbf{Nam Thoai:} Resources. \textbf{Vu Thi Hanh Thu:} Supervision, Writing - review, Funding acquisition, Visualization, Investigation \textbf{Nguyen Tuan Hung:} Supervision, Writing - review, Validation, Methodology, Investigation, Formal analysis.

\section*{Declaration of competing interest}
The authors declare that they have no known competing financial interests or personal relationships that could have appeared to influence the work reported in this paper.

\section*{Acknowledgements}
This work was financially supported by Vietnam National University Ho Chi Minh City (NCM2024-50-01).

\appendix
\section{Supplementary material}
Table of structural parameters and band gap; table of Born charges, Phonon frequency, and dielectric properties; figure of HSE06-calculated band structure; figure of electrostatic potential along vacuum axis ($z$); figure of electrostatic potential differences across interface for different alignment types; figure of band alignments for water splitting at pH 0, 2, and 7; figure of PBE band structure for the effective mass calculation; figure of total energy-strain relationship in zigzag and armchair directions; figure of CBM and VBM energies under uniaxial strain; figure of Phonon dispersion and DOS.

\section{\label{sec:A1}Linearized Boltzmann transport equation}
Our starting point is the linearized Boltzmann transport equation (BTE) for the non‐equilibrium carrier distribution \(f(\bm r,\bm k)\) under a weak and homogeneous electric field \(\bm{E}\). In a first‐principles framework, one solves the BTE to obtain the steady‐state deviation \(\Delta f = f - f^0\) from the equilibrium Fermi–Dirac distribution \(f^0_{n\bm k}\). The macroscopic mobility tensor \(\mu_{\alpha\beta}\) of the sample is then given by a spatial average of the local, space‐resolved tensor \(\mu_{\alpha\beta}(\bm r)\):
\begin{equation}\label{eq:mu_space}
\mu_{\alpha\beta}
\;=\;
\frac{1}{V} \int_V d^3r\;\mu_{\alpha\beta}(\bm r),
\end{equation}
where $V$ is the total volume of the system and $\mu(\bm r)$ is the space‐resolved mobility determined from the BTE:
\begin{equation}\label{eq:mu_local}
\mu_{\alpha\beta}(\bm r)
=
\frac{q}{n_c\,\Omega_{\rm uc}}
\sum_{n}
\int_{\rm BZ}
\frac{d^3 k}{\Omega_{\rm BZ}}\;
\bm{v}_{n\bm k}^{\alpha}\;\frac{\partial \Delta f_{n}(\bm r,\bm k)}{\partial E_\beta}\,.
\end{equation}
where $\alpha$ and $\beta$ are direction indices, $q$ is the carrier charge, $\Omega_{\rm uc}$ is the unit‐cell volume, $\Omega_{\rm BZ}$ is the Brillouin‐zone volume, \(\bm{v}_{n\bm k}^{\alpha} = (1/{\hbar})\times(\,\partial \varepsilon_{n\bm k}/\partial k_\alpha)\) is the band‐velocity of band $n$ at wavevector $\bm k$, $n_c$ is the carrier concentrations, which is defined for the electron and hole, respectively, as follows:
\begin{equation}\label{eq:ne}
n_e = \frac{1}{(2\pi)^3}\sum_{n}\int_{\rm BZ} f_{n\bm k}\,d^3k,
\end{equation}
and
\begin{equation}\label{eq:nh}
n_h = \frac{1}{(2\pi)^3}\sum_{n}\int_{\rm BZ}1 - f_{n\bm k}\,d^3k,
\end{equation}
where $f_n(\bm r,\bm k)$ is the non‐equilibrium distribution obtained by solving the linearized BTE, which is given by
\begin{equation}\label{eq:bte_lin1}
\bm{v}_{n\bm{k}}\cdot\nabla_{\bm r}f_{n\bm{k}}
+\frac{q}{\hbar}\,\bm{E}\cdot\nabla_{\bm k}f_{n\bm{k}}
=\left.\frac{\partial f}{\partial t}\right|_{\rm scattering}.
\end{equation}

Writing the linear‐response ansatz of the distribution function $f_{n\bm{k}}$ = $f^{0}_{n\bm{k}} + \Delta f_{n\bm{k}}$ and keeping only first‐order terms in \(\Delta f\) and \(\bm E\), one obtains:
\begin{equation}\label{eq:bte_lin2}
\bm{v}_{n\bm{k}}\cdot\nabla_{\bm r}\Delta f_{n\bm{k}}
+\frac{q}{\hbar}\,\bm{E}\cdot\nabla_{\bm k}f^0_{n\bm{k}}
= -\sum_{n'\bm{k}'}T_{n\bm{k},\,n'\bm{k}'}\,\bigl[\Delta f_{n\bm{k}} - \Delta f_{n'\bm{k}'}\bigr],
\end{equation}
where the total scattering operator $T_{n\bm{k},\,n'\bm{k}'}$ is expressed by both electron–phonon (e–p) and electron–defect (e–d) contributions as follows:
\begin{equation}
T_{n\bm{k},\,n'\bm{k}'}
= T^{\rm ph}_{n\bm{k},\,n'\bm{k}'}
+ T^{\rm d}_{n\bm{k},\,n'\bm{k}'}.
\end{equation}

\section{\label{sec:A2}The calculation method for the electron-phonon interaction}
We denote $T_{n\bm{k},\,n'\bm{k}'}$ is the transition rate for an electron scattering from state \((n,\bm{k})\) to \((n',\bm{k}')\). This rate comprises both electron–phonon and electron–defect contributions. In this subsection, we focus on the electron–phonon term; the electron–defect expression is given in Section B. Using Fermi’s golden rule, the phonon-mediated transition rate for absorption (\(+\)) or emission (\(-\)) of a phonon \((\nu,\bm{q})\) is:  
\begin{align}
T^{\pm,\nu\bm q}_{n\bm{k},n'\bm{k}'}
&= \frac{2\pi}{\hbar}
\frac{1}{N_k}
\bigl|g^{\nu}_{n\bm{k},\,n'\bm{k}'}\bigr|^2
\begin{cases}
n_{\nu\bm q}\,\delta\bigl(\Delta\varepsilon+\hbar\omega_{\nu\bm q}\bigr),\\
(1+n_{\nu\bm q})\,\delta\bigl(\Delta\varepsilon-\hbar\omega_{\nu\bm q}\bigr),
\end{cases}
\label{eq:T_ph_branch}
\end{align}
where \(N_k\) denotes the total number of \(\bm k\)-points used in the Brillouin‐zone sampling, 
\(\bm q = \bm k' - \bm k\) is the phonon wavevector, and \(\nu\) labels the phonon mode. 

We define
\[
\Delta\varepsilon \;\equiv\; \varepsilon_{n\bm k} - \varepsilon_{n'\bm k'}, 
\qquad
n_{\nu\bm q} = \frac{1}{e^{\hbar\omega_{\nu\bm q}/k_B T} - 1},
\]
where \(n_{\nu\bm q}\) is the Bose–Einstein occupation factor of phonons. The electron–phonon coupling matrix element is \(g^{\nu}_{n\bm k,\,n'\bm k'}\).

Summing over all phonon branches and both absorption (\(+\)) and emission (\(-\)) processes yields:
\begin{equation}\label{eq:T_ph_total}
T^{\rm ph}_{n\bm k,\,n'\bm k'}
= \sum_{\nu,\bm q}
\bigl[T^{+,\nu\bm q}_{n\bm k,\,n'\bm k'} 
     + T^{-,\nu\bm q}_{n\bm k,\,n'\bm k'}\bigr].
\end{equation}

The $\tau^{0}_{n\bm{k}}$ is self-energy relaxation time for electronic state $n\bm{k}$ then canculated using the fomular: $1/\tau^{\mathrm{ph,}0}_{n\bm{k}} = \sum_{n'\bm{k'}}\bar T_{n\bm{k},n'\bm{k'}}$ and $\bar T$ is the out-scattering part. 
From \(T^{\rm ph}\), the state‐dependent out‐scattering rate is:
\[
\frac1{\tau^{\rm ph,0}_{n\bm{k}}}
= \sum_{n'\bm{k}'} 
\frac{1 - f^0_{n'}(\bm{k}')}{1 - f^0_{n}(\bm{k})}\;T^{\rm ph}_{n\bm{k},\,n'\bm{k}'},
\]

By applying the relaxation time approximation method (RTA), then we have:
\begin{equation*}
    1/\tau^{m}_{n\mathrm{k}}=\sum_{n'\bm{k'}}\bar T_{n\bm{k},n'\bm{k'}},
\end{equation*} and consequently the $\tau^{\mathrm{ph},m}_{n\mathrm{k}} = \tau^{\mathrm{ph},0}_{n\mathrm{k}}$, which is call self-energy relaxation time approximation (SERTA).

Therefore, the phonon-induced scattering relaxation time \(\tau^{\mathrm{ph}}_{n\bm{k}}\) can be written explicitly as:

\begin{equation}
\begin{multlined}
\frac{1}{\tau^{\mathrm{ph}}_{n\bm{k}}}
 = \frac{2\pi}{\hbar} \frac{1}{N_k}
 \sum_{n'\bm{k}'} \bigl|g^{\nu}_{n\bm{k},\,n'\bm{k}'}\bigr|^2 \times \Bigl[
 (n_{\nu\bm q}+f^{0}_{n'}(\bm{k'}))\delta(\Delta\varepsilon+\hbar\omega_{\nu\bm q})
 \\ +(1+n_{\nu\bm q}-f^{0}_{n'}(\bm{k'}))\delta(\Delta\varepsilon-\hbar\omega_{\nu\bm q})
 \Bigr]
\end{multlined}
\end{equation}

The term $g^{\nu}_{n\bm{k},\,n'\bm{k+q}}(\bm{k,q})$ EPC matrix element is defined as:
\begin{equation}
    g^{\nu}_{n\bm{k},\,n'\bm{k+q}}(\bm{k,q})=\bigl\langle \psi_{n',\bm{k}+\bm{q}}\bigm|V_{\nu,\bm{q}}\bigm|\psi_{n,\bm{k}}\bigr\rangle, 
\end{equation}
where $V_{\nu \bm{q}}$ denotes the perturbing potential, which can be obtained via density functional perturbation theory (DFPT).

\section{\label{sec:A3}The calculation method for the electron-defect interaction}
For elastic scattering off static point defects (\(\hbar\omega=0\), no phonon occupation), Fermi’s golden rule gives for each “branch” (\(+\), absorption; \(-\), emission)
\begin{equation}\label{eq:T-ed}
T^{\pm,\rm d}_{n\bm{k},\,n'\bm{k}'}
= \frac{2\pi}{\hbar}\,
\frac{n_{\rm at}\,C_d}{N_k}\,
\bigl|M_{n\bm{k},\,n'\bm{k}'}\bigr|^2\,
\delta\bigl(\varepsilon_{n\bm{k}}-\varepsilon_{n'\bm{k}'}\bigr),
\end{equation}
where $n_{at}$ is the number of atoms in a  primitive cell, $C_d$ the (dimensionless) defect atomic concentration,

Since the process is elastic, thefore $T^{+,d}=T^{-,d}\equiv T^d$, and furthermore \(f^0_{n}(\bm k)=f^0_{n'}(\bm k')\) on the energy‐conserving shell.  Hence \(T_{n\bm{k},\,n'\bm{k}'}\) yields:
\begin{eqnarray}
T_{n\bm{k},\,n'\bm{k}'}
= \bar T_{n\bm{k},\,n'\bm{k}'}
= \widetilde T_{n\bm{k},\,n'\bm{k}'} \nonumber
\\= \frac{2\pi\,n_{\rm d}\,}{\hbar\,N_k}
\bigl|M_{n\bm{k},\,n'\bm{k}'}\bigr|^2
\,\delta\bigl(\varepsilon_{n\bm{k}}-\varepsilon_{n'\bm{k}'}\bigr),
\end{eqnarray}
Therefore, we obtain the equation for the inverse relaxation time
\begin{equation}
    \tau_{n\bm{k}}^{-1} = \frac{2\pi}{\hbar} \frac{n_d}{\Omega_{BZ}} \sum_{n \prime \bm{k}\prime} \int \mathrm{d}\bm{q} \left|M_{ n\prime\bm{k},n \bm{k}}\right|^2 \delta\left(\varepsilon_{n\prime\bm{k}}-\varepsilon_{n \bm{k}}\right)
\end{equation}
The term $M_{ n\prime\bm{k},n \bm{k}}$ is the e-d matrix elements they encode the probability amplitude for scattering from the unperturbed state $|n\bm{k}\bigl\rangle$ to due $|n'\bm{k'}\bigl\rangle$ to the perturbation potential $V_{d}$ from defect:
\begin{equation}\label{eq:M_def}
M_{ n\prime\bm{k},n \bm{k}} =\bigl\langle n'\bm k'\bigl|\Delta V_{d}\bigr|n\bm k\bigl\rangle, \quad \Delta V_{e\!-\!d}=V^{(d)}_{\rm KS}-V^{(p)}_{\rm KS}.
\end{equation}
use superscripts (d) and (p) to denote the defect-containing and pristine systems, respectively. When using, as we do here, norm-conserving pseudopotentials in the Kleinman-Bylander form, the Kohn-Sham potential can be written as a sum of local and nonlocal parts:
\begin{eqnarray}\label{eq:KBform}
    V_{KS}= V_L(\bm{r})+\hat{V}_{NL}\\ \nonumber
    V_L(\bm r)=V_H(\bm r)+V_{XC}(\bm r)+V_{pp}(\bm r).
\end{eqnarray}
We the yielding the separated the matrix elements into a local and a nonlocal part:
\begin{equation}
    M_{n'\bm k',\,n\bm k}
= M^{\rm L}_{n'\bm k',\,n\bm k}
+ M^{\rm NL}_{n'\bm k',\,n\bm k}.
\end{equation}
The local part in Fourier form is:
\begin{equation}\label{eq:ML}
M^{\rm L}_{n'\bm k',\,n\bm k}
=\sum_{\bm G}
\tilde V_L(\bm k'-\bm k+\bm G)\,
\bigl\langle u_{n'\bm k'}\bigr|e^{i\bm G\cdot\bm r}\bigl|u_{n\bm k}\bigl\rangle_{\rm uc}.
\end{equation}
Here, \(u_{n\bm k}(\bm r)\) denotes the cell-periodic part of the Bloch wavefunction, normalized over a primitive cell of volume \(\Omega_{\rm uc}\).  The vectors \(\bm G\) are the reciprocal-lattice vectors of the primitive cell.  The Fourier components of the local perturbation potential \(\Delta V_{\rm L}(\bm r)\), computed in a supercell of volume \(\Omega_{\rm sup}\), are defined by
\begin{equation}\label{eq:DVq}
\Delta\widetilde V_{\rm L}(\bm q)
= \frac{1}{\Omega_{\rm uc}}
\int_{\Omega_{\rm sup}} \!d^3r\;
\Delta V_{\rm L}(\bm r)\,e^{-i\bm q\cdot\bm r}\,,
\end{equation}
where the momentum transfer:
\[
\bm q = \bm k' - \bm k + \bm G
\]
appears in the scattering process \(\lvert n\bm k\rangle \to \lvert n'\bm k'\rangle\).
\\
The nonlocal matrix elements \(M^{\rm NL}_{n'\bm k,n\bm k}\) are defined as the difference between the nonlocal pseudopotential operators in the defect‐containing and pristine supercells:
\begin{eqnarray}\label{eq:MNL_def}
M^{\rm NL}_{n'\bm k,n\bm k}
= \langle n'\bm k|\hat V_{\rm NL}^{\rm d} - \hat V_{\rm NL}^{\rm p}|n\bm k\rangle
\nonumber\\= \langle n'\bm k|V_{\rm NL}^{\rm d}|n\bm k\rangle
- \langle n'\bm k|V_{\rm NL}^{\rm p}|n\bm k\rangle.
\end{eqnarray}
Each supercell nonlocal potential \(\hat V_{\rm NL}^\alpha\) (\(\alpha=\mathrm{d,p}\)) can be expanded in the Kleinman–Bylander projectors \(\beta_i^{(s)}\) for each species \(s\):
\begin{equation}\label{eq:VNL_KB}
\langle n'\bm k|\hat V_{\rm NL}^\alpha|n\bm k\rangle
= \sum_{s=1}^{n_s}\sum_{i,j}
D_{ij}^{(s),\alpha}(\bm G)\,
\langle n'\bm k|\beta_i^{(s)}\rangle\,
\langle \beta_j^{(s)}|n\bm k\rangle,
\end{equation}
where \(D_{ij}^{(s),\alpha}(\bm G)\) are the supercell‐Fourier coefficients of the KB projectors and \(\bm G\) is the reciprocal‐lattice vector consistent with \(\bm k'-\bm k\).  (See the derivation in Ref.~\cite{Lu2019-ch}.)

\section*{Data availability}
Data will be made available on request.
\bibliographystyle{elsarticle-num-names} 



\end{document}